

\input macro1.tex
\tolerance=100000
\hyphenpenalty=1000
\raggedbottom

\def\ainit{\hoffset=.0 truecm
            \voffset=1. truecm
            \hsize=17. truecm
            \vsize=23.5 truecm
            \baselineskip=11.pt
            \lineskip=0pt
            \lineskiplimit=0pt}
\def\pag{\pageno=2\footline={\hss\tenrm\folio\hss}}
\ainit
%
%

\def\mic{\,\mu{\rm m}}

\def\Msol{\,{\rm m_\odot}}
\def\Lsol{\,{\rm L_\odot}}
\def\sol{\,{\rm \odot}}
\def\Msyr{\,{{\rm m_\odot}\,{\rm yr}^{-1}}}
%
%

\def\lsim{\,\lower2truept\hbox{${< \atop\hbox{\raise4truept\hbox{$\sim$}}}$}\,}
\def\gsim{\,\lower2truept\hbox{${> \atop\hbox{\raise4truept\hbox{$\sim$}}}$}\,}

%
%
\def\oneskip{\vskip\baselineskip}
\centerline{\null}
\nopagenumbers
\oneskip
\noindent
\centerline{MODELS FOR THE EVOLUTION OF THE SPECTRAL ENERGY DISTRIBUTION}

\centerline{OF ELLIPTICAL GALAXIES FROM UV TO FAR--IR WAVELENGTHS}

\oneskip
\oneskip
\parindent=1truecm
\parskip 0pt

\centerline{P. MAZZEI\footnote{$^1$}{e-mail: mazzei@astrpd.astro.it
{\it or} 39003::mazzei}, and G. DE ZOTTI\footnote{$^2$}{e-mail:
dezotti@astrpd.astro.it {\it or} 39003::dezotti}}

\centerline{Osservatorio Astronomico, Vicolo dell'Osservatorio 5,
 I--35122 Padova, Italy}
\oneskip
\centerline{and}

\oneskip
\centerline{C. XU}

\centerline{Max-Planck-Institut f\"{u}r Kernphysik,
Postfach 103980, D 6900 Heidelberg, Germany}

\vfill\eject

\pag
\noindent
\centerline{ABSTRACT}

\smallskip

We have worked out evolutionary synthesis models of the broad-band
spectral energy distribution of elliptical galaxies
over the whole frequency range from UV to far--IR.
Internal extinction and far--IR re--emission by interstellar dust have been
taken into account in a self--consistent way.
Diffuse dust emission has been modelled in terms of two components: warm dust,
located in regions of high radiation intensity, and cold dust, heated by the
general radiation field. Emission from circumstellar dust clouds at different
galactic ages was also taken into account.
The models reproduce well the present  average broad--band spectrum of nearby
ellipticals over about four decades in frequency.
Under the assumption of a
dust--to--gas ratio proportional to the metallicity, the fraction of
bolometric luminosity coming out at far-IR wavelengths strongly evolves during
the galaxy lifetime, ranging from very low local values ($\lsim 0.5\%$)
to $\sim 30\%$ or more in the first billion years.
Some models even imply that early phases are optically thick; these models
provide an excellent fit of the observed spectral energy distribution of
the high redshift galaxy IRAS F10214$+$4724.
Far--IR observations may thus play a key role in investigating the
early evolution of elliptical galaxies. A strong
far-IR evolution of early-type galaxies might be a crucial ingredient to
explain the deep $60\mic$ IRAS counts.

\medskip\noindent
{\it Subject headings:} galaxies:evolution --- galaxies: photometry --
galaxies: stellar content --- galaxies: interstellar medium ---
interstellar: grains

\vfill\eject

\noindent
\centerline{1.\  INTRODUCTION}

\medskip
For many years optical surveys have been the primary tool for studying the
evolution of the population of galaxies. CCD cameras have allowed to
reach $B \simeq 27$, i.e. surface densities
as high as $\simeq 3\times 10^5\,\hbox{deg}^{-2}$
(Tyson 1988; Lilly, Cowie, \& Gardner 1991).
Counts and colors indicated a substantial increase with lookback time of
the surface density of blue galaxies, consistent with evolutionary stellar
population synthesis models (Guiderdoni \& Rocca-Volmerange 1990),
provided that the universe has a low density ($q_0 \simeq 0.05$) and galaxies
form at high redshifts ($z_{\hbox{for}} \simeq 30$).
However, the observed redshift distributions of galaxies down to $B \simeq 22$
(Broadhurst, Ellis, \& Shanks 1988; Koo 1988; Colless et al. 1990; Metcalfe
et al. 1991) or
down to $B \simeq 24$ (Lilly et al. 1991) cannot be accounted for
by pure luminosity evolution models. Various explanations
have been proposed, resorting to either density evolution
(Rocca-Volmerange \& Guiderdoni 1990; Cowie 1991) or to luminosity dependent
evolution (Broadhurst et al. 1988), or to a non-zero
cosmological constant $\Lambda$ (Lilly et al. 1991; Fukugita et al. 1990).

A great observational stride occurred thanks to the impressively
fast advances in infrared array technology, which have already made
possible to reach a limiting magnitude $K = 23$ (Cowie et al. 1990;
Cowie 1991), i.e. surface  densities $\simeq 2\times 10^5\,\hbox{deg}^{-2}$
close to those of the deepest optical surveys. Unlike optical counts, which
are highly
responsive to young stellar populations, infrared counts
are much less sensitive to evolution effects, and, hence,
are well suited for investigating the geometry of the universe.
The extremely interesting new result was
that counts in the K band show a remarkable flattening below $K \simeq 20$,
most easily explained in a flat $q_0=0.5$ universe with $\Lambda = 0$
(Cowie 1991).

Yet another, complementary,  perspective is offered by
IRAS counts (Hacking \& Houck 1987; Lonsdale et al. 1990),
which also show evidence of galaxy evolution (Hacking et al. 1987;
Danese et al. 1987; Lonsdale \& Hacking 1989): they provide
information on the dust content (and, hence, on one side on metal abundance,
on the other side on internal extinction) as a function of the galactic age.

Additional pieces of information are coming from
VLA surveys at $\mu$Jy levels, which may already see spiral disks
at redshifts of order unity (Fomalont et al. 1991). The tight correlation
between radio and far-IR luminosities of disc galaxies (Helou, Soifer, \&
Rowan-Robinson 1985; de~Jong et al. 1985; Wunderlich \& Klein 1988)
allows to set stringent constraints on models fitting IRAS data
(Franceschini et al. 1988).

It is clear from the above that in order to achieve a real progress
in understanding the history of galaxies as a whole it is crucial
to build consistent models able to exploit the observational data from all
wavebands. Our group has undertaken a long term program in this
direction. Mazzei, Xu, \& De Zotti (1992) have
carried out an exploratory study of the evolution of the continuum
spectrum of disc galaxies over the whole frequency range from UV to
far-IR. These galaxies correspond to the extreme case of a dissipational
collapse, where the gas depletion is slow, i.e. the star formation rate
(SFR) was never much higher than it is today
(Sandage 1986). At the other extreme, elliptical galaxies are thought
to have used up most of their gas to form stars in a time short compared
with the collapse time, i.e. their initial SFR should have been
spectacularly large. In this paper we investigate the evolution of the
broad-band properties of these galaxies. Franceschini et al. (1991)
have taken up preliminary results of Mazzei et al. (1992)
on the evolution of disk galaxies and of
the present paper to predict source counts and contributions to the
extragalactic background from near-IR to mm wavelengths. Chokshi et al.
(1993) have exploited some of the present models as well as those
by Mazzei et al. (1992) to work out simulations of deep optical and
near infrared extragalactic counts.

Several detailed stellar population synthesis models have been
published in recent years (Bruzual 1983; Arimoto \& Yoshii 1986;
Guiderdoni \& Rocca--Volmerange 1987; Buzzoni 1989; Brocato {\it et
al.} 1990; Pickles \& van der Kruit 1990; Charlot \& Bruzual 1991;
Bruzual \& Charlot 1993). They mostly cover the
wavelength range from UV to optical, and in all cases do not extend
beyond $\lambda = 3\,\mu$m.

On the contrary,
our aim is to investigate possible evolutionary behaviours of the overall
spectra of elliptical galaxies, from UV to far-IR, with emphasis on
the essentially unexplored long wavelength region. Given the exploratory
nature of this work, we are not particularly interested in a good
spectral resolution.

In addition to covering a much broader frequency range (albeit with a
coarse resolution), the present models
overcome several limitations affecting, to a different degree, previous
studies:

\item{\phantom{ii}i)} In the optical and near-IR regions we exploit
the recent sets of isochrones constructed by Bertelli et al.
(1990), which incorporate the results of
stellar evolutionary calculations for two values of gas
metallicity, $Z=0.001$ and $0.02$, have a fine coverage of masses
and ages, and include almost all evolutionary phases from the main
sequence to the stage of planetary nebula ejection or of carbon ignition,
as appropriate given the initial mass. Most of existing models, on the
other hand, use libraries of stellar evolutionary tracks which are
either incomplete or are assembled tying together tracks computed by
different authors, using different assumptions.

\item{\phantom{i}ii)} In the present study, photometric evolution is linked
(albeit in a rough manner, cf. \S 3.1)
to chemical evolution, while most previous studies do not take into
account the increased metallicity of successive stellar generations.
Of course, both stellar evolution and dust content of galaxies (and, hence,
internal extinction) are sensitive to the metal abundance and
consideration of chemical evolution sets important constraints on
crucial ingredients such as the time dependent star formation rate and the
initial mass function.

\item{iii)} We attempt to model in a self-consistent way
the evolution of internal extinction as a function of the metallicity
and of the mass in gas, taking into account not only the mean observed
colors of galaxies but also their dust emission properties, based
on IRAS data.

\medskip\noindent
The plan of the paper is the following.
In Sect. 2 we describe the adopted model for the chemical evolution of
galaxies. In Sect. 3 we derive the integrated spectrum from stellar
photospheres for several galactic ages. The emissions from hot
circumstellar dust clouds and from diffuse dust are discussed in Sections
4 and 5, respectively. Our main results, presented in Sect. 6, are discussed
in Sect. 7 and our conclusions are summarized in Sect. 8.

\bigskip

\centerline{2.\  CHEMICAL EVOLUTION}

\medskip
We have adopted Schmidt's (1959) parametrization, wherein the
star--formation rate (SFR), $\psi (t)$, is proportional
to some power, $n$, of
the fractional mass of gas in the galaxy, $f_g = m_{gas}/m_{gal}$,
assumed to be, initially, unity:
$$\psi (t) = \psi _0 f_g^n\ \Msyr . \eqno(1)$$
Current analyses favour $n = 1$--$2$ (see Buat 1992); in addition to these
values, we have also explored the case $n=0.5$, advocated by Madore (1977).
Successful models for early type galaxies require $\psi_0 \geq 100\,\Msyr$ for
$n \geq 1$ ($\psi_0 \geq 35\,\Msyr$ for $n = 0.5$).

As for the initial mass function (IMF), $\phi (m)$,
we have considered both Salpeter's (1955) form:
$$\phi (m) dm = A \left({m\over \Msol}\right)^{-2.35} d\left({m\over
\Msol}\right)\qquad m_l \leq m \leq m_u, \eqno(2)$$
with $m_u=100\,\Msol$ and $m_l$ ranging from $0.01\,\Msol$ to $0.1\,\Msol$,
and Scalo's (1986) form:
$$\phi (m) dm = A \cases {\left({m/ \Msol}\right)^{-1.25} d\left({m/
\Msol}\right)\quad &if\quad $m_l \leq m \leq 1\,\Msol$,\cr
\left({m/ \Msol}\right)^{-2.35} d\left({m/
\Msol}\right)\quad &if\quad $1\,\Msol < m \leq 2\,\Msol$,\cr
2^{0.35} \left({m/\Msol}\right)^{-2.7} d\left({m/
\Msol}\right)\quad &if\quad $2\,\Msol < m \leq m_u$,\cr}\eqno(3)$$
with $m_u=100\,\Msol$ and $m_l=0.1\,\Msol$. In the case of a Scalo
IMF the results are only very weakly dependent on the values of
$m_l$ and $m_u$.

The coefficient $A$ is fixed by the normalization condition
$$\int_{m_l}^{m_u} m \phi (m) dm = 1 , \eqno(4)$$
giving $A = 7.27\times 10^{-2}$ or $A = 0.172$ for a Salpeter
IMF and $m_l=0.01\,\Msol$ or $m_l=0.1\,\Msol$, respectively;
$A = 0.366$ for a Scalo IMF\null.

The galaxy is assumed to be a close system, i.e. we neglect both winds
and inflow of intergalactic gas. Supernova driven galactic winds
may well be important during the early evolutionary stages of ellipticals,
particularly for lower mass objects (Brocato et al. 1990). On the
other hand, the extended hot coronae around these galaxies,
indicated by X-ray observations (e.g. Trinchieri \& Fabbiano 1985), may
imply the existence of massive halos, capable of hampering or even of
preventing steady galactic winds, or of accretion flows.
Matteucci (1992) found that, in the presence of a dynamically
dominant dark matter component distributed like stars, ellipticals
develop winds very late;
Worthey, Faber, \& Gonzales (1992) pointed out that the observed  trend
of the $[Mg/Fe]$ ratio in elliptical galaxies is at odds with
the wind hypothesis.
In any case, a reliable modelling of these effects is very difficult
(see also Ciotti et al. 1991).

Stellar lifetimes are taken into account i.e. we do not resort to the
instantaneous recycling approximation. The gas is assumed to be well mixed
and uniformly distributed.

The variations with galactic age, T, of the fractional gas mass $f_g(T)$
[and, through eq. (1), of the SFR, $\psi(T)$] and of the gas
metallicity $Z_g(T)$ are obtained by numerically solving the standard equations
for chemical evolution (Tinsley 1980; Arimoto \& Yoshii
1986; for details see Mazzei et al. 1992).

Examples of our results, for $m_g(0) = m_{gal} = 10^{11}\,\Msol$,
$\psi(0) = 100\,\Msol\,\hbox{yr}^{-1}$ and different choices for the
shape of the IMF, for the
lower stellar mass limit $m_l$, and for the
power law index $n$ of the SFR [eq.~(1)] are shown in Fig. 1. Panels $b)$ and
$d)$ show the evolution with the galactic age
of the ``mean stellar metallicity'' $\langle Z_s(T)\rangle$,
weighted by the luminosities in the $V$ band, $L_V$:
$$\langle Z_s(T)\rangle= {{\int_0^T\psi(T-\tau)L_V(\tau)Z_g(T-\tau)\,d\tau}
\over{ \int_0^T\psi(T-\tau)L_V(\tau)\,d\tau}}, \eqno(5)$$
where $\tau$ is the age of a given stellar generation.

The gas depletion rate [see Fig.~1, panels $a)$ and $c)$] is obviously
faster
if the power law index $n$ of the assumed SFR [eq.~(1)] is significantly
flatter than unity, and is correspondingly slower if $n > 1$. If
$n = 0.5$, the mean metal abundance of stars is maximum at $T \simeq 2\,$Gyr
(panel $d)$ of Fig.~1) and slightly decreases afterwards due to the
dilution effect of mass loss from older, metal poorer, stars.

For given SFR, a Salpeter IMF with $m_l = 0.01\,\Msol$
implies a faster decrease of the gas fraction (because low mass stars
do not contribute to the gas supply through mass loss) than for larger
$m_l$ values or for a Scalo IMF. A Salpeter IMF implies both a
slower  metal enrichment of the gas and a higher mass to light ratio
because a larger mass fraction is locked up in low mass stars which do not
contribute to the chemical enrichment and emit little light.

At $T=15\,$Gyr, for $\psi_0 = 100\,\Msyr$,
a Salpeter IMF yields a mass to light
ratio $\simeq 30\,\Msol/ \Lsol$; this value decreases slightly with
increasing $\psi_0$ down to $\simeq 20\,\Msol/ \Lsol$.
At $10\,$Gyr, for the same $\psi_0$ we find $M/L \simeq 25\,\Msol/\Lsol$.
Increasing the lower mass limit, $m_l$, to 0.05 or to
0.1 $\Msol$ the mass to light ratio at $15\,$Gyr decreases to
 $15$ or $11\,\Msol/ \Lsol$, respectively; at 10 Gyr to 13 or
$6.7\,\Msol/ \Lsol$. For a Scalo IMF, the
mass to light ratio at 15 Gyr is $4.5\,\Msol/ \Lsol$; at 10 Gyr it becomes
$2.4\,\Msol/ \Lsol$.

All models predict higher values for $Z_g$  than for coeval disc
systems (see Mazzei et al. 1992). For a Salpeter IMF
the final $\langle Z_s\rangle$ is close to
the solar value for $m_l=0.01$ $\Msol$, about twice solar for $m_l=0.05$
$\Msol$
and $\approx 0.1$ for $m_l=0.1$ $\Msol$; the last value of $\langle Z_s\rangle$
also applies to a Scalo IMF (Fig. 1).

For comparison, Munn's (1992) data, extending out to 0.5--1
half--light radii, $r_e$, for seven early type galaxies
indicate metallicities ranging from 2--3 times
solar in the centers, to solar at $1\,r_e$.

The final gas fraction is higher for higher $m_l$ and for a Scalo IMF.

\bigskip\bigskip
\centerline{3.\  PHOTOMETRIC EVOLUTION}

\medskip
\centerline{3.1.\ {\it Synthetic starlight spectrum}}

\medskip\noindent
The synthetic spectrum of stellar populations as a function of the
galactic age was derived from $UV$ to the
$N$ band ($10.2$ $\mic$) using the method
described by Mazzei et al. (1992). We recall here only the main points.

The contribution of a stellar generation of age $\tau$
to the integrated luminosity in
the passband $\Delta \lambda$ is given by:
$$ l_{\Delta \lambda} (\tau) = \int _{m_{min}}^{m_{max}(\tau)} \phi (m)
10^{-0.4 (M_{\Delta \lambda} (m, \tau) - M_\odot)} dm\ \ \Lsol \Msol^{\! -1},
\eqno(6)$$
where $m$ is the initial stellar mass,
$m_{min}$ is the minimum mass represented in the isochrone,
$m_{max}(\tau)$ is the maximum mass of stars still visible
at the age $\tau$, i.e. the
largest mass which has not yet reached  the stage of either the final
explosion or of the formation of a collapsed remnant, $M_{\Delta
\lambda}(m,\tau)$
is the absolute magnitude of a star of initial mass $m$ and age $\tau$, and
$M_\odot = 4.72$ is the bolometric luminosity of the sun.

The global luminosity at the galactic age $T$ is then obtained as the sum
of the contributions of all earlier generations, weighted by the appropriate
SFR:
$$ L_{\Delta \lambda} (T) = \int _{0}^{T} \psi (T - \tau) l_{\Delta \lambda}
(\tau)\, d\tau\ \ \Lsol . \eqno(7)$$
The number of stars born at each galactic age $T$ and their metallicity
are obtained by solving the equations governing the chemical evolution,
with the SFR and IMF specified above.

To describe their distribution in the H--R diagram we have adopted,
as already mentioned, the theoretical isochrones derived by Bertelli et
al.~(1990) for metallicities Z=0.001 and Z=0.02, extended by Mazzei (1988)
up to $100\Msol$ and to an age of $10^6\,$yr.

Luminosities and effective temperatures of stars born with metallicities
$0.001 < Z(t) < 0.02$ have been computed with a simple interpolation,
linear in $\log Z$,
between the values appropriate for the extremes. For $Z > 0.02$ we have
adopted the values corresponding to $Z = 0.02$.

The UBV (Johnson) magnitudes corresponding to
the luminosities and effective temperatures along the isochrones are
also taken from Bertelli et al. (1990). The red and infrared
magnitudes are obtained
using Johnson's (1965, 1966) colors for the various spectral types and
luminosity classes.
Monochromatic fluxes are computed using Johnson's (1966) calibrations.

The contributions to UV luminosities of MS stars with $T_{eff} \geq 5500\,$K
are estimated exploiting Kurucz's (1979) models with $\log g = 4$
($g$ being the surface gravity in cgs units). For
stars with lower effective temperatures  we have used data from
the IUE Ultraviolet Spectral Atlas (Wu et al. 1983).

The computed (unreddened) broad--band galaxy spectra
for galactic ages $T = 2$, 5, 10 and $15\,$Gyr,
for a model with $\psi_0=100$ $\Msyr$ and $n=1$,
are displayed in Fig.~2 (heavy line). This
figure shows that the resulting spectral energy distribution
(SED) is largely independent of the lower mass limit, $m_l$.
Some differences arise only in the UV region in advanced
phases of galaxy evolution ($T > 5 \,$ Gyr) owing to a higher fraction of
residual gas (see Fig.~1). Clearly, the earliest stages of galaxy evolution
($T < 5$ Gyr) are the most
sensitive to the {\it shape} of the IMF and of the SFR (see Fig.~3).
At later times, the UV emission depends primarily on the residual
gas fraction (and is thus higher for higher values of $n$) and
on the mass fraction not locked up in low mass stars.
Optical--IR
colors, on the other hand, are essentially identical, for $T > 5\,$Gyr,
in all cases, not surprisingly since the bulk of the star formation activity
occurs anyway in the first few Gyr.

\medskip
\centerline{3.2.\ {\it Corrections for internal extinction}}

\medskip\noindent
We assume elliptical galaxies to be spherically symmetric with well-mixed
dust and stars. Their stellar density distribution is conveniently described
by King's (1962) formula (radii are in units of the ``core radius'' $r_c$):
$$\rho \propto ( 1+r^2)^{-3/2}, \eqno(8)$$
up to a cut-off radius $\log r_t = 2.2$ (King 1966).

The surface brightness at a projected distance $s$ from the galactic
center is:
$$I_{\lambda} (s) = \int_{-z_{\rm max}}^{+z_{\rm max}} dz\,\rho_{\lambda} (r)
\exp\left[-\tau_{\lambda}(s,z)\right], \eqno(9)$$
where $r=(s^2+z^2)^{1/2}$ and
$$\tau_{\lambda} (s,z) = \tau_{0_{\lambda}} \int_{z}^{z_{\rm max}} \,dz'\,
\left({\rho(r')\over \rho_0}\right)_{\hbox{dust}},\eqno(10)$$
with $r'=(s^2+z'^2)^{1/2}$.

Following Guiderdoni \& Rocca-Volmerange (1987) we assume a dust to gas
ratio proportional to some power of the metallicity, so that
$$ \tau_{\lambda}(T) = \tau_{0_{\lambda}}\left({Z_g (T)\over Z_g(15) }\right)^s
\left ({f_g(T) \over f_g(15)}\right) ,\eqno(11)$$
with
$$\tau_{0_{\lambda}}= \tau_\star
\left( {A_{\lambda}\over A_V}\right) Z_{\sol} \left(
{Z_g(15)\over Z_{\sol}}\right)^s f_g(15) \, \eqno(12)$$
and $s = 1.6$ for $\lambda > 2000\,$\AA and $s = 1.35$ for $\lambda
< 2000\,$\AA. We have adopted the  interstellar extinction curve given by
Seaton
(1979) for $\lambda \le 0.37\mic$ and by Rieke \& Lebofsky (1985) at
longer wavelengths. The index $15$ refers to a galactic age, $T$, of 15 Gyr.

The coefficient $\tau_\star$ was derived from the mean far-IR to optical
luminosity ratio of elliptical galaxies, assuming that all
the absorbed starlight is eventually re-emitted by dust in the far-IR.

IRAS data for a large
sample of early-type galaxies were reported by Knapp et al. (1989).
{}From these data, Mazzei \& De Zotti (1993) have selected  a complete sample
of
elliptical galaxies and derived their average far-IR properties exploiting
survival analysis  techniques.
They find $L_{FIR}/L_B\approx 1\times10^{-2}$, where
$L_B=(\lambda f_{\lambda})_B$ and $L_{FIR}$ is the
total dust luminosity between $42.5$ $\mic$ and $125.5 \mic$.

\bigskip
\centerline{4.\  EMISSION FROM CIRCUMSTELLAR DUST}

\medskip\noindent
As discussed by Ghosh, Drapatz, \& Peppel (1986) and Cox, Kr\"ugel,
\& Metger (1986) the main contribution from
stellar sources at mid--IR wavelengths and beyond comes
from evolved objects with circumstellar dust shells, such as
OH/IR stars.
The observed color temperatures of the dust clouds enshrouding these stars
are $\gsim 200\,$--$\, 500\,$K, so
that most of the emission comes out between 5 and $20\mic$.

OH/IR stars are believed to be in the final stage of evolution along the
asymptotic giant branch (AGB). Their pulsation and kinematic properties are
consistent
with initial masses in the range $m_{HeF} \leq m \leq m_{up}$
($m_{HeF}$ being the maximum mass for
violent ignition of helium burning and $m_{up}$ the  maximum
mass of stars that develop a highly degenerate CO core), i.e. in the range
where stars experience a prolonged, thermally pulsating AGB phase (Engels
et al. 1983; Baud et al. 1981; Feast 1963).

We assume the contribution of OH/IR stars to the emission of a galaxy
at each galactic age $T$ to be a fixed fraction $F$ of the global
bolometric luminosity of AGB stars with initial masses in the quoted range.
The spectrum of OH~27.2+0.2 (Baud et al. 1985)
was assumed to be representative for stars of this class (see also Cox \&
Mezger 1989). Then, the total luminosity of OH/IR stars in the passband
$\Delta\lambda$ is given by:
$$ L_{OH,\Delta\lambda} (T) = F \int _{T_{AGB}(m_{up})}^T d\tau\,\psi(T-\tau)
\int _{m_{l,OH}(\tau)}^{m_{u,OH}(\tau)} \phi (m)
10^{-0.4 (M_{\Delta \lambda} (m, \tau) - M_\odot)} dm\ \ \Lsol , \eqno(13)$$
where ${T_{AGB}(m_{up})}$ is the time when the first OH/IR
stars appear, $m_{l,OH}(\tau)\ (\geq m_{HeF})$ and
$m_{u,OH}(\tau)\  (\leq m_{up})$ are the minimum and
the maximum mass of OH/IR stars of age $\tau$.
The coefficient $F$ was determined by Mazzei et al. (1992)
from the condition that
OH/IR stars account for 10\% of the observed $12\mic$ luminosity of our
Galaxy (Ghosh et al. 1986; Boulanger \& P\'erault 1988). They find
$F=0.05$, in good agreement with Herman \& Habing's (1985) estimate
based on a different approach.

\bigskip
\centerline{5.\  DIFFUSE DUST EMISSION}

\medskip\noindent
The diffuse dust emission spectrum is modelled
following Xu \& De~Zotti (1989), i.e. taking into account
the contributions of two components: warm dust, located in regions
of high radiation intensity (e.g., in the neighborhood of OB clusters)
and cold dust, heated by the general interstellar radiation field.
The model allows for a realistic grain--size distribution and includes
PAH molecules. The emission spectrum of PAH's is computed following
Puget, L\'eger, \& Boulanger (1985) and using the absorption cross sections
and the bandwidths at 3.3, 6.2, 7.7, 8.6 and $11.3\mic$ given
in Table 2 of Puget \& L\'eger (1989).
We have computed the  contributions of PAH in the IRAS
$12\mic$ band which encompasses the emission features at $7.7\mic$,
$8.6\mic$, and $11.3\mic$, and in the ISOCAM filters LW2 and SW1,
covering the wavelength ranges 5--$8.5\mic$ and 3.05--$4.1\mic$,
respectively.

The temperature distribution of cold dust at each galactocentric radius
is determined by the local intensity of the interstellar radiation
field, assumed to be described by a King's (1962) formula
(see \S3.2.). Following Jura (1982) the {\it core radius}, $r_c$, is taken
to be $100\,$pc
and the central radiation field intensity, $I_c$, is
$16\,\hbox{V~mag}\,\hbox{arcsec}^{-2}$ corresponding to $I_0=46 I_
{\rm local}$, $I_{\rm local}$ being the radiation field intensity in the
solar vicinity (see Xu \& De Zotti 1989). The cold dust is assumed to be
uniformly distributed up to a cutoff radius, $r_d$. Since the dust
temperature drops quickly with increasing galactocentric radius following
the decrease in the radiation intensity, its emission spectrum
peaks at longer and longer wavelengths as $r_d$ increases.
For example, if $r_d=30r_c$ it peaks around $100\mic$  while for
 $r_d=100 r_c$ it peaks around  $150\mic$.

The radiation intensity in starforming regions, $I_w$, determining the warm
dust temperature distribution, and the warm to cold luminosity ratio,
$R_{w/c}$,
were derived from a fit of the observed average ratios $f_{25\mic}/f_{60\mic}$
and $f_{60\mic}/f_{100\mic}$ derived by Mazzei \& De Zotti (1993) for a
complete sample of nearby ellipticals. We find $I_w=110\,I_{\rm local}$
and $R_{w/c} = 0.53$ if $r_d=30r_c$; $I_w=90\,I_{\rm local}$ and $R_{w/c} =
0.27$ if $r_d=100r_c$.

The relative contributions of the warm and cold components are evolving with
galactic age; we assume $R_{w/c}$ to be proportional to the star formation
rate. The amount of starlight absorbed and re--emitted by dust
is determined at each time using the model described in {\S}3.2.

\bigskip
\centerline{6.\  RESULTS}

\medskip\noindent
At least three factors are playing a key role in regulating the
evolution of the spectral energy distributions of galaxies: the global stellar
birthrate, the initial mass function, and dust extinction. Since
the present understanding
of the physical mechanisms which determine the gas/star conversion rate is
very poor, and the information on the star formation history is rudimentary,
we have, as usual, parametrized our ignorance by adopting simple laws,
consistent with the available data, and have explored the effect of
varying the relevant parameters with the aim of offering a comprehensive
scenario of possibilities. In our models, dust extinction is simply related
to the gas fraction and to the metallicity, i.e. is determined by
the SFR and the IMF. For some choices of the parameters, early type galaxies
are predicted to be opaque, i.e. to emit most of their bolometric luminosity
in the far-IR. For broader ranges of the parameter values, however,
we predict rather blue colors during early phases. Both kinds of objects
may have been observed. Analyses of implications of different models
for counts and redshift distributions of faint galaxies in different
wavebands will then help in understanding how galaxies really evolve and
in clarifying the driving factors.

\medskip
\centerline{6.1.\ {\it Spectral distribution of starlight}}
\smallskip

In Fig. 4
the spectral distributions of starlight at a galactic age of 15 Gyr
for different choices of the SFR and of the IMF are shown
and compared with data on nearby ellipticals. The predicted spectra are only
weakly dependent on galactic age if the latter varies between 10 and 15 Gyr.
As is well known, a fit
of the optical to near-IR continuum spectrum is easily obtained with
a fast decreasing SFR (timescale $\lsim 1\,$Gyr; Guiderdoni \& Rocca-Volmerange
1987; Bruzual \& Charlot 1993). In fact, the galactic luminosity in this
spectral region is dominated by red giant stars with an age of several
Gyr, so that it is only weakly sensitive to the model parameters
such as the IMF, $\psi_0$ (provided that it is large enough), $n$.

On the other hand, ellipticals display a
considerable variety of far-UV properties (the far-UV to optical luminosity
ratio varies by a factor of 10). In the present framework the UV emission
comes from main sequence stars, as a consequence of the residual star
formation; a relatively high far-UV branch is thus more easily
accounted for with a Scalo IMF, implying a higher residual gas
fraction (see Fig.~1), hence a higher residual SFR [after eq.(1)].
For a Scalo IMF and $n=1$ the data imply
$\psi_0 \geq 100\Msyr$; the far-UV luminosity decreases slightly for higher
values of $\psi_0$ or lower values of $n$.
The coldest UV spectra can be accounted for with a Salpeter IMF, $m_l =
0.01\Msol$, $n = 0.5$ and $\psi_0 \simeq 100\Msyr$. For such IMF, the UV
branch rises with increasing $n$; a less conspicuous increase is obtained
increasing $m_l$.

To explain the far-UV branch of ellipticals,
several authors have rather advocated post-AGB stars (Barbaro \& Olivi 1989;
Bertelli, Chiosi, \& Bertola 1989), or post-RGB stars (Greggio \& Renzini
1990), or
post-EAGB stars (Brocato et al. 1990). A thorough discussion of this
problem is outside the scope of the present
paper since the spectral distribution of starlight
is primarily exploited here to deal in a self consistent way with the
radiation field heating the dust. Given the small energetic content of the
UV branch, a refined modelling is only of marginal importance for our purposes.

As for the evolution with galactic age, we may note that going backwards
from 15 Gyr to 2 Gyr, the bolometric luminosity increases by a factor
of about 10. The evolution during earlier phases is of course heavily
dependent on the choice of $\psi_0$ and $n$.

On the other hand, the {\it observed} optical/near-IR (in the
rest frame of the galaxy) luminosity during the early evolutionary
phases is strongly dependent on the amount of internal extinction.
The model described in \S 3.2. implies that about 25--30\% of starlight is
absorbed by dust at $T=2\,$Gyr if $n=1$ and for a Salpeter IMF. For the
same IMF and $n=0.5$ the absorbed fraction is substantially larger:
at $T=2\,$Gyr the galaxy is opaque; a $\simeq 30\%$ absorption occurs at
$T=5\,$Gyr. The dust absorption is expected to be much weaker
in the case of a Scalo IMF: the absorbed fraction is only 4--5\% at 2 Gyr.

\medskip
\centerline{6.2.\ {\it Circumstellar dust emission}}
\smallskip

The contribution from circumstellar dust shells
to the galaxy light, illustrated by Figs. 2 and 3,
is higher during the first billion years of the galaxy lifetime;
the duration of this phase is shorter for higher values of
$\psi_0$ and, given $\psi_0$, for lower $n$ and/or  lower $m_l$.
The energy radiated by circumstellar dust at $T=2$ Gyr in the range
5--$20\,\mu$m is always
twice  the pure stellar emission in the same wavelength range; it
amounts to 1--2$\%$  of the total galaxy luminosity.

For a Scalo IMF, for which our model yields a moderate evolution of
interstellar dust emission with galactic age, circumstellar
dust provides at least 50$\%$ of the $12 \mic$ galaxy luminosity at
any galactic age. A somewhat lower fraction (30\%)
obtains in the
case of a Salpeter IMF with $n = 1$ if $m_l = 0.1\Msol$; if $m_l =
0.01\Msol$, the contribution of circumstellar dust to the $12\mic$
luminosity is $\simeq 30$--40\% in the first several Gyr of the
galactic lifetime, but drops to $\simeq 20\%$ after 10 Gyr.

\medskip
\centerline{6.3.\ {\it Diffuse dust emission}}
\smallskip

There is a growing evidence that ellipticals have a significant
interstellar medium, and that some star formation may take place
(Bally \& Thronson 1989; Thronson, Bally, \& Hacking 1989;
Knapp et al. 1989; Lees et al.
1991; Roberts et al. 1991; Bregman, Hogg, \& Roberts 1992).

As already mentioned, in the present framework the UV luminosity comes from
hot main sequence stars. In particular, at an age of 15 Gyr, assumed to be
typical for present day ellipticals, in the case of a Salpeter IMF
OB stars account for about 1\% of the bolometric luminosity
if $m_l = 0.01\Msol$ or for about 6\% if $m_l = 0.1\Msol$;
for a Scalo IMF their contribution is about 4\%.

According to Leisawitz
\& Hauser (1988)  roughly 37\% of the total radiated energy from an
OB cluster is absorbed by nearby dust grains, and should therefore
be re-radiated in the form of ``warm'' dust emission. We then expect
that the latter corresponds to 0.3--2\% of the bolometric luminosity
of elliptical galaxies.
A fit of the mean IRAS colors of nearby elliptical galaxies (see \S 5)
implies that warm dust accounts
for about \frac1/3 of the dust emission, i.e. for about 0.1--0.3\% of the
bolometric luminosity of these galaxies, not far from the above
expectation.
So far as Leisawitz \& Hauser's estimate holds, this might be
taken as evidence in favour of some star formation in ellipticals; on
the other hand, a substantial UV excess (corresponding to several percent
of the galactic luminosity) could be more easily interpreted in terms
of hot evolved stars.

As already mentioned, in the case of a Salpeter IMF the models
predict a strong increase of dust absorption with decreasing
galactic age. As a consequence, the ratio of far-IR to bolometric
luminosity strongly evolves during the galaxy lifetime;
for the model in Fig.~5 it increases from
$0.4\%$--$0.75\%$ at $T = 15\,$Gyr, to
$2\%$--$3\%$ at $T = 10\,$Gyr, to
$13\%$--$18\%$ at $T = 5\,$Gyr, and to
$26\%$--$31\%$ at $T = 2\,$Gyr, the lower values corresponding to
$r_d=30r_c$, the upper values to $r_d=100r_c$.

For $n=0.5$ and a Salpeter IMF, elliptical galaxies during
the earliest phases of their evolution radiate mostly in the far-IR;
for the model in Fig. 6, the far-IR to bolometric luminosity ratio
increases from 0.5\% at $T= 15\,$Gyr, to 4\% at $T= 10\,$Gyr,
26\% at $T= 5\,$Gyr, 91\% at $T= 2\,$Gyr.

The relative amounts of the warm and cold dust components are also
evolving with galactic age. Since their ratio is
assumed to be proportional to the star formation rate,
the contribution of the warm component to $L_{\rm dust}$
generally increases from
$21\%$--$35\%$ at $T = 15\,$Gyr (for $r_d=100~r_c$ and $30r_c$, respectively),
to $50\%$--$66\%$~, $91\%$--$95\%$, and $99\%$ at
$T = 10\,$Gyr, $5\,$Gyr and $2\,$Gyr, respectively (see Fig.~4).

\bigskip
\centerline{7.\ DISCUSSION}

\medskip\noindent
The aim of this Section is to focus the main effects of different choices
for the SFR and for the IMF, as well as of the geometry of the dust
distribution. It must be kept in mind, however, that
there are also considerable uncertainties in stellar evolution theory,
which may significantly, or even strongly, affect the conclusions. A
particularly debated issue is the effect of overshooting, discussed
in \S7.4.

\medskip
\centerline{7.1.\ {\it Effect of the SFR}}
\smallskip

The influence of different choices of the power law index $n$ [eq.~(1)]
has been discussed
by Mazzei (1988). Gas poor systems are more easily explained
with low values of $n$, implying stronger evolution
as shown by Fig.~6. A rapid gas consumption,
coupled with the model for dust absorption described in \S 3.2.,
entails an optically--thick phase
during the first $2\, $Gyr of evolution.

Higher values of $\psi_0$ obviously imply a faster gas consumption and
a correspondingly faster dimming of the galaxy (see Fig.~7).
For example, if $n=1$, models
with $\psi_0 = 1000\Msyr$ keep brighter than models with $\psi_0 = 100\Msyr$
for $T \lsim 1\,$Gyr; during later phases
the dimming of former relative to the latter is, however, modest
at all wavelengths except for the UV where the emission directly
reflects the SFR.

\medskip
\centerline{7.2.\ {\it Initial Mass Function}}
\smallskip

As already mentioned, a Scalo IMF yields a higher UV branch and a slower
evolution in the far-IR in comparison
with a Salpeter IMF. This follows, in the present framework, from the
slower gas depletion rate in the Scalo case, as illustrated by Fig. 1.
In turn, this is due to the fact that the Salpeter IMF is steeper
below $1\Msol$ so that a higher mass fraction is locked up in stars
which do not have significant mass loss.

In Salpeter case, while the resulting SED's are quite insensitive to
the assumed lower mass limit (due to the negligible contribution of
low mass stars to the galactic luminosity,
cf. Fig.~8), different values of $m_l$
entail significant differences in the chemical evolution, again because
low mass stars do not contribute to the chemical enrichment (see Fig. 1).

\medskip
\centerline{7.3.\ {\it Geometry of the dust distribution}}
\smallskip

Our assumption of a uniform dust distribution within a radius $r_d$ is
clearly an oversimplification. The dust is probably clumpy, and there
may be relatively dust free lines of sight towards some actively star
forming regions. As a consequence, even galaxies with a very high far-IR
to optical luminosity ratio and, hence, very high {\it average} dust
extinction, may appear quite blue. Colors predicted by the present
models in the optical band are therefore particularly uncertain and may
be biased towards the red.

\medskip
\centerline{7.4.\ {\it Effect of overshooting}}

\smallskip
The most important effect of overshooting, in the present framework,
is the decrease of the critical masses $m_{up}$ and $m_{HeF}$ which
determine at which age a galaxy develops for the first time AGB and
RGB stars, respectively (cf. Renzini \& Buzzoni 1986; Chiosi, Bertelli,
\& Bressan 1988).

Although, if overshooting is effective, $t_{AGB}$ is a factor of several
larger than predicted by classical stellar evolution models, this
critical epoch still occurs too early ($t_{AGB} \simeq 10^8\,$yr)
for the corresponding changes in the galactic colors to be detectable.
On the other hand, convective overshooting increases $t_{RGB}$ from
$\simeq 1\, $Gyr to $\simeq 2.5\, $Gyr, thus making it easier
to detect the associated increase in the near--IR luminosity. Hence
ISOCAM observations of high redshift galaxies may also provide a test
for overshooting.

\bigskip
\centerline{8.\ CONCLUSIONS}

\medskip\noindent
We have presented the results of a first attempt to investigate in
a self--consistent way the evolution with cosmic time of the global
broad--band spectrum of elliptical galaxies from UV to far--IR wavelengths.
Our analysis incorporates up--to--date results on stellar evolution theory
and allows for absorption and re--emission of starlight by dust.
Models reproduce well the average broad--band spectrum of nearby ellipticals
over a factor of $10^4$ in wavelength.

We have centered this paper on the essentially unexplored long wavelength
range. Our results show that IR observations may provide important
clues to understand the evolution of elliptical galaxies.

The contribution of circumstellar dust surrounding OH/IR stars to
the $12\mic$ luminosity of early type galaxies
of age $T$ is, in the present framework, directly related to the SFR
a few Gyr earlier, since most of it comes from stars
whose lifetimes is $\sim 1$--2 Gyr. Correspondingly, for local
galaxies ($T\sim 10$--$15\,$Gyr), models entailing a small residual
star formation (such as those with a Salpeter IMF and $m_l = 0.01\Msol$)
predict that OH/IR stars contribute only $\simeq 5\%$
of the $12\mic$ luminosity of early type galaxies. Such fraction
increases to 40--50\% for models with a Scalo IMF. For comparison,
Knapp, Gunn, \& Wynn-Williams (1992) estimated that about 40\% of the
observed $12\mic$ emission of normal ellipticals is probably due
to circumstellar dust.

The total mid-IR luminosity of OH/IR stars increases with decreasing
galactic age, following the increase of the SFR. Hence mid-IR observations
would be informative on this crucial quantity.

If, as assumed here, the dust content is proportional to the gas fraction
and to the metallicity, the very large initial star formation rates
(which ensure a prompt metal enrichment) of spheroidal galaxies imply
that they can be very dusty during the
first few Gyr of their evolution, in agreement with the arguments put forward
by van den Berg (1990). For a broad range of parameter values,
during these evolutionary
phases they are expected to emit in the far-IR
$\sim 30\%$ of their bolometric luminosity, i.e an amount comparable to
their visible/near-IR luminosity. Some systems may even become optically thick,
as illustrated by the case $n=0.5$ in Fig. 6 and in Table 1;
the spectral energy distribution predicted by this model for a
galactic age $T\simeq 1\,$Gyr matches remarkably well that of
the ultraluminous galaxy IRAS F10214$+$4724 (Rowan-Robinson et al. 1991;
Lawrence et al. 1993) from UV to sub-mm wavelengths (see Fig. 9).

Our results also suggest that spheroidal galaxies might undergo a substantial
evolution of the 60--$100\mic$ color, at variance with disc galaxies for
which it is expected not to vary much, at least for $T > 2\,$Gyr. In
fact, for discs Mazzei et al. (1992) found that
$\log(F_{60\mic}/F_{100\mic})$ increases only from $\simeq -0.6$ at 15 Gyr
to $\simeq -0.4$ at 2 Gyr. On the other hand, we expect that the warm dust
component becomes increasingly important for higher and higher SFR and
eventually dominates dust emission during the earliest evolutionary phases.
Correspondingly, $\log(F_{60\mic}/F_{100\mic})$ should increase from
$\simeq -0.5$ at 15 Gyr to $\simeq +0.2$ at 2 Gyr.

The strong far-IR luminosity evolution of early type galaxies
predicted by the present models, at variance with the case of disc
systems whose far-IR emission is expected to keep essentially constant
at least for $T > 2\,$Gyr (Mazzei et al. 1992), might be an essential
ingredient to explain the deep $60\mic$ counts
(Hacking \& Houck 1987; Lonsdale et al. 1990).

\bigskip\noindent
{\it Acknowledgements.} We thank the Referee, dr. Perry Hacking, for
his constructive comments. Work supported in part by CNR, ASI and by the
EEC program Human Capital and Mobility.

\def\aa #1 #2{{A\&A,}~{ #1}, {#2}}
\def\aar #1 #2{{A\&A Rev,}~{ #1}, {#2}}
\def\aas #1 #2{{A\&AS,}~{ #1}, {#2}}
\def\araa #1 #2{{ARA\&A,}~{ #1}, {#2}}
\def\aj #1 #2{{AJ,}~{ #1}, {#2}}
\def\alett #1 #2{{Ap Lett,}~{ #1}, {#2}}
\def\apj #1 #2{{ApJ,}~{ #1}, {#2}}
\def\apjs #1 #2{{ApJS,}~{ #1}, {#2}}
\def\ass #1 #2{{Ap\&SS,}~{ #1}, {#2}}
\def\baas #1 #2{{BAAS,}~{ #1}, {#2}}
\def\jrasc #1 #2{{JRASC,}~{ #1}, {#2}}
\def\mmras #1 #2{{MmRAS,}~{ #1}, {#2}}
\def\mnras #1 #2{{MNRAS,}~{ #1}, {#2}}
\def\nat #1 #2{{ Nature,}~{ #1}, {#2}}
\def\pasj #1 #2{{PASJ,}~{ #1}, {#2}}
\def\pasp #1 #2{{PASP,}~{ #1}, {#2}}
\def\qjras #1 #2{{QJRAS,}~{ #1}, {#2}}

\def\physrep #1 #2{{Phys.Rep.,}~{ #1}, #2}








\def\ref{\noindent\hangindent=20pt\hangafter=1}

\oneskip
\centerline {\bf References}
\oneskip
\parindent=0pt
\parskip=0pt

\ref
Arimoto, N., \& Yoshii, Y. 1986, \aa 164 260

\ref
Bally, J., \& Thronson, H.A. 1989, \aj 97 69

\ref
Barbaro, G.,\&  Olivi, F.M. 1989, \apj 337 125


\ref
Baud, B., Habing, H.J., Matthews, H.E., \&  Winnberg, A. 1981, \aa 95 156

\ref
Baud, B., Sargent, A.J., Werner, M.W., \&  Bentley, A.F. 1985, \apj 292 628

\ref
Bertelli, G., Chiosi, C., \& Bertola, F. 1989, \apj 339 903

\ref
Bertelli, G., Betto, R., Bressan, A., Chiosi, C., Nasi, E., \&
Vallenari, A. 1990, \aas 85 845

%
%

\ref
Boulanger, F., \& P\'erault, M. 1988, \apj  330 964

\ref
Bregman, J.N., Hogg, D.E., \& Roberts, M.S. 1992, \apj 387 484

\ref
Broadhurst, T.J., Ellis, R.S., \& Shanks, T. 1988, \mnras 235 827

\ref
Brocato E., Matteucci, F., Mazzitelli, I., \& Tornamb\'e A. 1990, \apj 349 458

\ref
Bruzual, A.G. 1983, \apj  273 105

\ref
Bruzual, A.G., \& Charlot, S. 1993, \apj 405 538

\ref
Buat, V. 1992, \aa 264 444

\ref
Burstein, D., Bertola, F., Buson, L.M., Faber, S.M., \& Lauer, T.R. 1988, \apj
328 440

%

\ref
Charlot, S., \& Bruzual, G.A. 1991, \apj 367 126

\ref
Chiosi, C., Bertelli, G., \& Bressan, A. 1988, \aa  196 84

\ref
Chokshi, A., Lonsdale, C., Mazzei, P., \& De Zotti, G. 1993, ApJ, in press

\ref
Ciotti, L., D'Ercole, A., Pellegrini, S., \& Renzini, A. 1991, 376 380

\ref
Coleman, G.D., Wu, C.C., \& Weedman, D.W. 1980, \apjs 43 393


\ref
Colless, M.M., Ellis, R.S., Taylor, K. \& Hook, R.N., 1990, \mnras 244 408

\ref
Cowie, L.L., Gardner, J.P., Lilly, S.J., \& McLean, I. 1990, \apj  360 L1

\ref
Cowie, L.L. 1991, in {\it Relativistic Astrophysisc, Cosmology and Fundamental
Physics}, eds. Barrow, J.D., Mestel, L., Thomas, P.A., Annals of the New York
Accademy of Science, 247, p. 31

\ref
Cox, P., Kr\"ugel, E., \& Mezger, P.G. 1986, \aa  155 380

\ref
Cox, P., \& Mezger, P.G. 1989, \aar 1 49

\ref
Danese, L., De Zotti, G., Franceschini, A., \& Toffolatti, L. 1987, \apj 318
L15

\ref
de~Jong, T., Klein, U., Wielebinski, R., \& Wunderlich, E. 1985, \aa 147 L6

\ref
Downes, D., Radford, S.J.E., Greve, A., \& Thum C. 1992, \apj 398 L25

\ref
Engels, D., Kreysa, E., Schultz, G.V., \& Sherwood, W.A. 1983, \aa 124 123

\ref
Feast, M.W. 1963, \mnras  125 367


\ref
Fomalont, E.B., Windhorst, R.A., Kristian, J.A. \& Kellermann, K.I. 1991,
\aj 102 1258

\ref
Franceschini, A., Danese, L., De~Zotti, G., \& Toffolatti, L. 1988,
\mnras 233 157

\ref
Franceschini, A., Toffolatti, L., Mazzei, P., Danese, L., \& De~Zotti, G. 1991,
 \aas 89 285

\ref
Fukugita, M., Takahara, F., Yamashita, K., \& Yoshii, Y. 1990, \apj 361 L1

\ref
Ghosh, S.K., Drapatz, S., \& Peppel, U.C. 1986, \aa  167 341

\ref
Glass, I.S. 1984, \mnras 211 461

\ref
Greggio, L., \& Renzini, A. 1990, \apj 364 35

\ref
Guiderdoni, B., \& Rocca--Volmerange, B. 1987, \aa  186 1


\ref
Guiderdoni, B., \& Rocca--Volmerange, B., 1990, \aa 227 362


\ref
Hacking, P., Condon, J.J. \& Houck, J.R. 1987, \apj 316 L15

\ref
Hacking, P. \& Houck, J.R. 1987, \apjs 63 311


\ref
Helou, G., Soifer, B.T. \& Rowan-Robinson, M. 1985, \apj 298 L7

\ref
Herman, J., \& Habing, H.J. 1985, \physrep 124 255


\ref
Impey, C.D., Wynn-Williams, C.G., \& Becklin, E.E. 1986, \apj 309 572

\ref
Johnson, H.L. 1965, \apj  141 170

\ref
Johnson, H.L. 1966, \araa  4 193

\ref
Jura, M. 1982 \apj 254 70

\ref
Kennicutt, R.C. jr. 1992, \apjs  79 255

\ref
King, I. 1962, \aj 67 471

\ref
King, I. 1966, \aj 71 64

\ref
Knapp, G.R., Guhathakurta, P., Kim, D.W., \& Jura, M. 1989, \apjs 70 257

\ref
Knapp, G.R., Gunn, J.E., \& Wynn-Williams, C.G. 1992, \apj 399 76

\ref
Koo, 1988, in {\it Towards Understanding Galaxies at Large Redshift},
ed. R.G. Kron, \& A. Renzini, A. (Dordrecht: Kluwer), 275

\ref
Kurucz, R.L. 1979, \aas 40 1

\ref
Lawrence, A., et al. 1993, \mnras 260 28

\ref
Lees, J.F., Knapp, G.R., Rupen, M.P., \& Phillips, T.G. 1991, \apj 379 177

\ref
Leisawitz, D., \& Hauser, M.G. 1988, \apj 332 954

\ref
Lilly, S.J., Cowie, L.L, \& Gardner, J.P. 1991, \apj 369 79



\ref
Lonsdale, C.J., \& Hacking, P.B. 1989, \apj 339 712

\ref
Lonsdale, C.J., Hacking, P.B., Conrow, T.P., \& Rowan-Robinson, M. 1990,
\apj 358 60

\ref
Madore, B.F. 1977, \mnras 178 1

\ref
Matteucci, F. 1992, \apj 397 32

\ref
Mazzei, P. 1988, Ph. D. thesis, Intern. School for Advanced Studies,
Trieste

\ref
Mazzei, P., \& De Zotti, G. 1993, ApJ, submitted

\ref
Mazzei, P., Xu, C., \& De Zotti, G. 1992, \aa  256 45

\ref
Metcalfe, N., Shanks, T., Fong, R., \& Jones, L.R. 1991, \mnras  249 498

\ref
Munn, J.A. 1992, \apj  399 444

%

\ref
Oke, J.B., \& Sandage, A. 1968, \apj  154 21



\ref
Puget, J.L., \& L\'eger, A. 1989, \araa  27 161

\ref
Puget, J.L., L\'eger, A., \& Boulanger, F. 1985, \aa  142 L19

\ref
Renzini, A., \& Buzzoni, A. 1986, in Spectral Evolution of Galaxies,
ed. C. Chiosi, \& A. Renzini (Dordrecht: Reidel), 195

\ref
Rieke, G.H., \& Lebofsky, M.J. 1985, \apj  288, 618

\ref
Roberts, M.S., Hogg, D.E., Bregman, J.N., Forman, W.R., \& Jones, C. 1991,
\apjs 75 751

\ref
Rocca-Volmerange, B. \& Guiderdoni, B. 1990, \mnras 247 166

\ref
Rowan-Robinson, M., et al. 1991, \nat 351 719

\ref
Rowan-Robinson, M., et al. 1993, \mnras 261 513

\ref
Salpeter, E.E. 1955, \apj  121 161

\ref
Sandage, A. 1986, \aa 161 89

\ref
Schild, R., \& Oke, J.B. 1971, \apj 169 209

\ref
Schmidt, M. 1959, \apj 129 243


\ref
Seaton, M.J. 1979, \mnras  187 73P

\ref
Scalo, J.M. 1986, \fcp 11 1

\ref
Telesco, C.M. 1993, \mnras 263 L37


\ref
Thronson, H.A., Bally, J., \& Hacking, P. 1989, \aj 97 363

\ref
Tinsley, B. 1980, \fcp 5 287

\ref
Trinchieri, G., \& Fabbiano, G. 1985, \apj 296 447

\ref
Tyson, J.A. 1988, \aj 96 1

\ref
van den Bergh, S. 1990, \pasp  102 503

%

\ref
Worthey, G., Faber, S.M., \& Gonzales, J.J. 1992 , \apj 398 69

\ref
Wu, C.--C., et al. 1983, NASA Newsletter No. 22

\ref
Wunderlich, E. \& Klein, U. 1988, \aa 206 47

\ref
Xu, C., \& De Zotti, G. 1989, \aa 225 12

\vfill\eject

\advancepageno
\centerline{\bf Figure captions}

\bigskip\noindent
{\bf Fig. 1.}  Gas fraction and mean stellar metallicity [eq.(5)]
as a function of the galactic age
for models with the same initial value of the SFR, $\psi_0=100\Msyr$.
Panels $a)$ and $b)$ refer  to $n=1$ [see eq.~(1)],
but different IMF and/or different lower mass limit $m_l$: the dot-dashed
and dotted lines corresponds to a Salpeter IMF with $m_l=0.05\Msol$ or
$m_l=0.1\Msol$, respectively; long dashes correspond to a
Scalo IMF with $m_l=0.1\Msol$.
Panels c) and d) correspond to a Salpeter IMF with a lower mass limit
$m_l=0.01\Msol$ but different values of $n$ [eq.~(1)]: $n=0.5$ (long
$+$ short dashes), $n=1$ (dots $+$ dashes), and $n=2$ (short dashes).

\bigskip\noindent
{\bf Fig. 2.} Broadband spectral distribution of stellar emission
(not corrected for dust extinction) at several galactic ages $T$
including (upper panel) or excluding
the contribution from circumstellar dust,
for two models, both with  $\psi_0=100\Msyr$, $n=1$, and a Salpeter IMF
but different mass limit:  $m_l=0.01\Msol$ (heavy lines), and
$m_l=0.1  \Msol$ (light lines).

\bigskip\noindent
{\bf Fig. 3.} Broadband spectral distribution of stellar emission
(not corrected for dust extinction) at several galactic ages $T$
including (upper panel) or excluding
the contribution from circumstellar dust,
for two models, both with  $\psi_0=100\Msyr$; heavy lines refer to a
Salpeter IMF with $m_l=0.01\Msol$, and $n=0.5$; light lines to
a Scalo IMF with $m_l=0.1 \Msol$ , and $n=1$.

\bigskip\noindent
{\bf Fig. 4.}
Broad--band spectra from UV to far-IR, corrected for internal extinction,
at $T=15\,$Gyr for models with different SFR's  and IMF's,
compared with data for nearby ellipticals (Burstein et al.
1988). The dashed line corresponds to a
Salpeter IMF with $m_l=0.01\Msol$, $\psi_0=100\Msyr$
and $n=0.5$; the dot dashed line to a Salpeter IMF with $m_l=0.1\Msol$,
$\psi_0=1000\Msyr$ and $n=1$; short $+$ long dashes to
a Scalo IMF with  $m_l=0.1\Msol$, $\psi_0=1000\,\Msyr$
and $n=1$. Open circles
refer to NGC 4649, a prototype UV-bright galaxy;
filled circles show the UV SEDs, normalized to the visible light,
of four early-type galaxies (N~4472, N~3115, N~4374, and N~1404) plus the
bulge of M~31, whose UV properties correspond to the sample average.
Open diamonds show the spectrum of ellipticals with low UV emission
(NGC 3379, NGC 4125, and NGC 4406).
Also plotted are data
by Schild \& Oke (1971), in the range from $0.34\mic$ to $0.89\mic$
(filled triangles),
recently confirmed  by Kennicutt's (1992)
large aperture observations of NGC~4472 and NGC~4486 in the range from
0.365 to $0.71\mic$, and the data by Oke and Sandage (1968) in the range
$0.7$ to $1.08\mic$ (stars).

\bigskip\noindent
{\bf Fig. 5.}
Broad--band spectrum from far--UV to far--IR
wavelengths, corrected for internal extinction,
at different galactic ages for a model with a Salpeter IMF,
$m_l=0.01\Msol$, $n=1$, and
 $\psi_0=100\Msyr$. Asterisks show the contribution
from circumstellar dust around OH/IR stars. The total emission of interstellar
dust (dotted line) is decomposed into the contributions of warm dust
(dashed line) and cold dust (dot--dashed line). Both dust components
include PAH molecules, which dominate the diffuse dust
emission at mid--IR wavelengths. The cold dust component is negligible
at $T=2\,$Gyr, so that the dashed line gives also the total diffuse dust
emission. The model spectrum at $T = 15\,$Gyr is compared with the
observed average SED of nearby ellipticals.
Data from $0.13\mic$ to $0.3\mic$ (filled circles)
are from Burstein et al.
(1988) and show to the UV SED's, normalized to the visible light,
of four early-type galaxies plus the
bulge of M~31 whose UV properties correspond to the mean of the
whole sample. Data from $0.36\mic$ to $1\mic$ (filled squares) are from
Coleman, Wu, \& Weedman (1980).
We have also adopted the following average near-IR colors:
$J-K=1.0\pm 0.1$, $V-K=3.2\pm 0.2$
(Glass 1984), and   $K-L=0.22 \pm 0.08$ (Impey et al. 1986).
The far--IR points correspond to the mean ratios $S_{12\mic}/S_B=0.97$,
$S_{25\mic}/S_B=0.5$, $S_{60\mic}/S_B=0.63$, $S_{100\mic}/S_B=1.78$
determined by Mazzei and De Zotti (1993) for a complete sample of
elliptical galaxies.

\bigskip\noindent
{\bf Fig. 6.} Effect of varying the index $n$ on the evolution with galactic
age, $T$, of the overall broad--band spectrum
of elliptical galaxies. The models correspond to a
Salpeter IMF with $m_l = 0.01\Msol$ and
$\psi_0 = 100\Msyr$, and two choices of the index $n$: $n=1$ (light lines)
and $n=0.5$ (heavy lines). The spectrum at $T=15\,$Gyr is compared with the
same data as in Fig. 4. The spectral energy distributions corresponding to
these two models at galactic ages of 2, 5, 10, and 15 Gyr are given in Table 1.

\bigskip\noindent
{\bf Fig. 7.} Effect of varying the initial SFR, $\psi_0$, on the
evolution with galactic age, $T$,
of the overall broad--band spectrum
of elliptical galaxies. Models assume a Scalo IMF with $m_l=0.1\Msol$,
$n=1$, and $\psi_0=1000\Msyr$ (light lines), or $\psi_0=100\Msyr$
(heavy lines). The spectrum at $T=15\,$Gyr is compared with the
same data as in Fig. 4.

\bigskip\noindent
{\bf Fig. 8.} Effect of varying the lower mass limit $m_l$ on the
evolution with galactic age, $T$,
of the overall broad--band spectrum
of elliptical galaxies. Models assume $\psi_0 = 1000\Msyr$, $n=1$, and
a Salpeter IMF with $m_l = 0.01\Msol$ (light lines) or
$m_l = 0.1\Msol$ (heavy lines).
The spectrum at $T=15\,$Gyr is compared with the
same data as in Fig. 4.

\bigskip\noindent
{\bf Fig. 9.} Spectral energy distribution predicted by one of the models
in Fig.~6 ($n=0.5$, Salpeter IMF with $m_l = 0.01\Msol$, $\psi_0 = 100\Msyr$),
for a galactic age $T=1\,$Gyr, shown in comparison with the observed spectrum
of the high redshift galaxy IRAS F10214$+$4724. The data, referred to
rest frame wavelengths, are from Rowan-Robinson et al. (1993),
Downes et al. (1992), and Telesco (1993).

\bye